\documentstyle[11pt]{article} 
\newcommand{\ba}{\begin{array}}
\newcommand{\ea}{\end{array}}
\newcommand{\bd}{\begin{displaymath}}
\newcommand{\ed}{\end{displaymath}}
\newcommand{\be}{\begin{equation}}
\newcommand{\ee}{\end{equation}}
\newcommand{\bea}{\begin{eqnarray}}
\newcommand{\eea}{\end{eqnarray}}

\def\a{\alpha}

\def\g{\gamma}

\def\L{\Lambda}

\def\q2 {q^2}
\def\sz {\sin^2\theta_W}

\def\r {\rightarrow}

\def\slash {\!\!\!\!\!\!/}

\begin{document}

\begin{flushright}
{\large MRI-PHY/96-36\\ December 1996\\ hep-ph/9612321}
\end{flushright}

\begin{center}
{\Large\bf
Probing Gauge-Mediated Supersymmetry Breaking Through Polarized 
Electron Beams in an $e^+e^-$ Collider}\\[20mm]
{\large{\sf Ambar Ghosal\footnote{Electronic address: ambar@mri.ernet.in}, 
Anirban Kundu\footnote{Electronic address: akundu@mri.ernet.in}}
and {\sf Biswarup Mukhopadhyaya\footnote{Electronic address: 
biswarup@mri.ernet.in}}}\\
{\em Mehta Research Institute,\\
10, Kasturba Gandhi Marg, Allahabad - 211 002, India}
\end{center}
\begin{abstract}

Using the facts that in Gauge-Mediated Supersymmetry Breaking schemes, masses 
of the right and the left sfermions can differ widely, and the gravitino is 
the Lightest Supersymmetric Particle, we show that it is possible to obtain 
unambiguous signatures of such schemes in a high energy $e^+e^-$ collider if 
one looks at the asymmetries in the cross-sections for certain final states 
with left-and right-polarized beams. 

\end{abstract}

\vskip 1 true cm

PACS numbers: 14.80.Ly, 13.85.Qk, 12.60.Jv

\newpage

While the search for supersymmetry (SUSY) as a fundamental symmetry of nature
continues to be an area of intense activity in high energy physics, it is not
clear that even the discovery of superpartners of the existing 
particles will answer a crucial question on the subject, namely how is
SUSY broken so that we have a consistent phenomenology. The popular paradigm,
as embodied in models based upon supergravity (SUGRA), is
that SUSY is broken in a `hidden sector' and the breaking is conveyed to the
observable sector through gravitational interaction, 
resulting in a gravitino
(the spin-3/2 partner of the graviton) with a mass of the order of the 
electroweak scale \cite{sugra}. In this picture, all
superparticle decays culminate into the production of the lightest 
supersymmetric particle (LSP) which is stable. In most models the LSP is a 
mixture of the
spin-1/2 superpartners of the photon, the Z-boson and the two neutral Higgs
bosons in the theory, and is designated as the lightest `neutralino'. 
Characteristic signals with such an `invisible' LSP have been widely 
explored \cite{lspsearch}.

Another scenario which has lately received a lot of attention 
is one where the Standard Model (SM) gauge group itself may act as the carrier
of SUSY breaking \cite{gmsb,cdfeegg,gmsbsigs,mass,babu}. 
For this, one requires a `messenger sector' which inherits the hidden sector 
effects through a separate set of interactions decoupled from the observable 
sector, at a scale that can be as low as ${\cal O}(10-100)$ TeV. 
The gauginos and sfermions acquire their masses through ordinary 
gauge interactions with the messenger sector at the 
one- and two-loop levels respectively \cite{gaume}.
This mechanism is referred to as Gauge Mediated SUSY Breaking
(GMSB). The recently renewed interest in GMSB is partially due to  the 
observation of a single
$e^+e^-\gamma\gamma + p_T\slash$ ~event at the Collider Detector at Fermilab
(CDF) experiment, which can be explained in this scenario \cite{cdfeegg}.
Another advantage of such a scenario is that flavour-diagonal sfermion
masses are induced at a rather low energy scale. Consequently, 
flavour changing neutral currents (FCNC) are naturally suppressed, which is 
a distinct advantage over the conventional SUGRA models.  

The messenger sector in GMSB models consists of lepton and quark superfields, 
and one (in extended GMSB models, more than one) superfield ($S$). 
The scalar as well as the auxiliary component of $S$ acquire finite
vacuum expectation values (VEV) when SUSY is broken in the 
messenger sector. 
The hidden sector and the interactions between the hidden 
and the messenger sector are responsible for giving VEVs to $S$. The 
messenger lepton and quark superfields must be in a GUT representation
(like a ${\boldmath 5}+{\boldmath {\overline 5}}$ or ${\boldmath 10}+
{\boldmath {\overline {10}}}$ of $SU(5)$ or $SO(10)$ respectively) to preserve 
the successful prediction of $\sz$.  

To find out from experiments which one is the real SUSY breaking scheme,
perhaps the most important thing to remember  is that the gravitino in GMSB 
is decoupled from the
SUSY breaking mechanism, and can be so light as to be considered massless
compared to the electroweak scale. Consequently, the gravitino becomes the LSP
now. The erstwhile LSP 
now becomes the next lightest SUSY particle (NLSP) which can decay to its
SM partner and the gravitino. This decay mode normally does not affect other
particles as the gravitational coupling is extremely weak.

While there have been several attempts in recent times  to
focus upon distinctive signals of GMSB \cite{gmsbsigs}, 
we want to point out here that a
very interesting way will be to explore the final states in a high energy
$e^{+}e^{-}$-annihilation experiment with a polarized electron beam. This
method utilizes the fact that the masses of left-and right handed sleptons
in GMSB are bound to be widely apart, as a consequence of their different 
$SU(2)\times U(1)$ quantum numbers which in turn dictate the gauge interactions
with the messenger sector.  This should be compared with SUGRA models where, 
except for the stau and the stop, all left-handed sfermions are nearly 
degenerate with the corresponding right-handed ones. The same principle 
sometimes causes a right-handed slepton to be the NLSP in GMSB, instead of 
a neutralino. We demostrate that in either
case one may expect to get widely asymmetric signals with polarized electron 
beams, which provide a clear distinction to the GMSB scenario. Here we 
demonstrate our predictions in the context of a linear collider with
$\sqrt{s} = 500 GeV$, because (i) the expected efficiency of electron
polarization is quite high ($\approx 90\%$) there, and (ii) such an energy
range covers a large part of the parameter space of a GMSB scenario.

We consider a minimal GMSB scenario where there is a single 
symmetry-breaking superfield $S$ in the messenger sector.
The masses induced for the gauginos ($M_{1/2}$) and the sfermions ($M_0$), 
induced at one-and 
two-loop levels respectively, depend crucially on two quantities. 
These are $M$, the messenger mass scale,
and $\Lambda$, the ratio between the VEV's of the auxiliary and scalar
components of $S$. The expressions for the induced masses are \cite{mass}

\bea
M_{{1}\over{2}}(M) &=& N_mf_1(\L/M){\a_i(M)\over 4\pi}\L,\\
M^2_0(M) &=& 2N_mf_2(\L/M)\sum_{i=1}^3k_iC_i\Big({\a_i(M)\over 4\pi}
\Big)^2\L^2,
\eea
with the messenger scale threshold functions
\bea
f_1(x)&=&{{1+x}\over x^2}\log(1+x)+(x\r -x),\\
f_2(x)&=&f_1(x)-{{2(1+x)}\over x^2}\Big[Li_2\Big({x\over 1+x}\Big)
-{1\over 4}Li_2\Big({2x\over 1+x}\Big)\Big]+(x\r -x).
\eea
In (1), $C_i=0$ for all gauge singlets and equals to $4/3$, $3/4$,
$(Y/2)^2$ for scalars falling in the fundamental representations of
$SU(3)$, $SU(2)$ and $U(1)$ respectively ($Y=2(Q-T_3)$ is the usual 
weak hypercharge), and $k_i=1,1,5/3$ for these three groups respectively
(our $\a_1$ is not GUT-normalized). $N_m$ is the number of messenger 
generations; for one pair of $5+\overline 5$, $N_m=1$, while for one pair
of $10+\overline{10}$, $N_m=3$ (these are the two values we will work
with). Once the mass terms
are obtained in this way, one can evolve them down to the electroweak scale.
Thus, all the slepton, squark chargino and neutralino  masses can be obtained 
from four inputs, viz. $M$, $\Lambda$, $\mu$ (the Higgsino mass parameter) and
$\tan\beta$ (the ratio of the two Higgs VEV's). The latter two
are treated as free parameters here. Of course, one has to 
diagonalize the chargino and neutralino mass matrices in order to get the 
physical masses for them. We have also taken care of the usual $D$-term and 
sfermion threshold corrections while evolving the sfermion mass
down to the electroweak scale. Equations (1) and (2) indicate that
in general the lightest neutralino is the NLSP for $N_m=1$ whereas for
higher $N_m$, the fact that $M_0 \sim N_m^{1/2}$ tends to make the 
right-handed sleptons the NLSP \cite{note1}. However, the mass spectrum also
depend crucially upon the diagonalisation of the neutralino mass matrix, and,
as is evident from the samples listed in Table 1, in some cases the
lightest neutralino is the NLSP even for $N_m = 3$.

Now, first consider the situation where the NLSP is a neutralino. The 
production of a pair of such NLSPs will be followed by each decaying into a 
photon and a 
gravitino leading to the signal $\gamma\gamma + p_T\slash $ . The analysis is
simpler if one assumes the NLSP to be Bino-dominated, which is indeed the
case for a large region of the parameter space. Using polarized
electron beams in a high-energy $e^{+}e^{-}$-collider, the rate for such a 
signal is not the same for left- and right electron beams. This is because 
neutralino pair production receives contributions from a t-channel diagram
mediated by a selectron. In this scheme, the right selectron (which is going to
be in the propagator when a right electron is involved) is normally much
lighter than a left selectron, as a result of which the t-channel contribution
is larger with a right-polarized electron beam. The s-channel contribution
on the other hand is nearly independent of electron polarization. At high 
energies ($\sim 500 GeV$) the t-channel contributions dominate over s-channel
unless the selectron is excessively heavy. Thus if one defines 
$\sigma^{\gamma\gamma}_{L(R)}$
as the cross section for 
$e^{+}e^{-} \longrightarrow \gamma \gamma + p_T\slash $ ~with a 
left(right)-polarized electron  and a specific $p_T\slash $ ~-cut, then 
$\sigma^{\gamma\gamma}_R$ is
bound to be larger than $\sigma^{\gamma\gamma}_L$. 
The effect is rather nicely described by
an asymmetry  parameter defined as

\be
{\cal{A}}^{\gamma\gamma} = 
{\frac{\sigma^{\gamma\gamma}_L - \sigma^{\gamma\gamma}_R}
{\sigma^{\gamma\gamma}_L + \sigma^{\gamma\gamma}_R}}
\ee

All systematic effects cancel out in the asymmetry ${\cal{A}}^{\gamma\gamma}$ 
which is plotted in Figure 1 for 
a centre-of-mass energy of 500 GeV and different 
values of the GMSB parameters. 
Note that such left-right asymmetries for exclusive final states can
be measured very precisely \cite{czar}.
Here we have applied a $p_T\slash $ ~-cut of
30 GeV for demonstration; the qualitative features do not depend qualitatively
on the cut. Basically, the quantity $\L$ controls the
masses induced in the observable sector; the dependence on M is only through
the way it affects the evolution of the coupling and mass parameters from SUSY
breaking scale in the messenger sector down to the electroweak scale. This 
latter dependence is relatively minor. On the other hand, a higher $\L$
is instrumental in causing a larger left-right mass splitting between 
selectrons, thereby leading to increasing negative values of
${\cal{A}}^{\gamma\gamma}$. 

The backgrounds come mainly from $\gamma \gamma Z$-production followed by
invisible decays of the Z. The total cross-section for such final states
is about $2 \times  10^{-3}$ pb, which is rather small for most of the
parameter range here. In addition, the backgrounds do not have any asymmetry
since they arise from chirality-preserving gauge interactions. Nonetheless,
they can reduce ${\cal{A}}^{\gamma\gamma}$ by enhancing the denominator if the
backgrounds are comparable to the signals. To alleviate such possibilities,
we show here a rather conservative scan over parameters, showing only those
cases where the signal is not less than $0.01$ pb. We find that sufficiently 
large asymmetries are still predicted over a wide region.

Consider next the other situation, where  right sleptons are the
NLSPs. This happens for most of the parameter space for $N_m=3$.
To be very precise, here the three right slepton generations are
lighter than any other superparticle and are 
practically degenarate unless one considers the fact that in the case of
a stau there is a non-negligible left-right mixing that makes the lighter 
stau the true NLSP. It is easy to see, however, that even then our arguments
go through when it comes to asymmetric production of right selectrons with
polarized electron beams. This is because, even though a stau may be lighter 
than it, a right selectron has no other way to decay but into an electron and
the gravitino. The consequent signal  from  a pair of right selectrons then is
$e^{+}e^{-} \r e^{+}e^{-} + p_T\slash $. Again, the rates
with a left-polarized electron beam is highly suppressed because the
t-channel becomes ineffective.  Thus one expects a large negative asymmetry
in $\sigma^{ee}_{L(R)}$, defined as the cross section for
$e^{+}e^{-} + p_T\slash $ ~with a left(right) electron beam, and with a suitable
$p_T\slash $ ~-cut so as to make the signal identifiable. The production of
left-handed selectrons cannot give rise to final states of this type.

There is however another point here. Unlike in the case with the $\g\g 
+p_T\slash $ ~final state, here we have a substantial contribution coming 
from W-pair production (the next largest background is from
Z-pairs and is down by roughly one order). This contribution receives an
enhancement with left-handed electrons because of t-channel effects. The
net effect is thus to cancel the asymmetry arising from GMSB, not
to speak of the relative suppression it causes to our signals in the total 
cross-section. What we have done, therefore, is to define the relevant 
asymmetry more carefully, in the following way:

\begin{equation}
{\cal{A}}^{ee} = \frac{\sigma^{tot}_L - \sigma^{tot}_R}
{\sigma^{tot}_L + \sigma^{tot}_R}
\end{equation}
where $\sigma^{tot}_{L(R)} = \sigma^{ee}_{L(R)} + \sigma^{WW}_{L(R)}$,
$\sigma^{WW}_{L(R)}$ being the contributions to the same final states
from a W-pair.

Plots of ${\cal A}^{ee}$, defined in this manner, are presented in Figure 2,
also drawn with a $p_T\slash $ ~-cut of 30 GeV.
Again, to avoid the error due to  Z-pair backgrounds etc., we have made the
rather conservative choice of only those regions in the parameter space 
where the SUSY cross-sections are at least 40$\%$ of the corresponding
W-induced rates. As is evident from the graphs, even then the asymmetry
is quite spectacular. Starting from large negative values, ${\cal{A}}^{ee}$
gradually becomes positive as the selectron masse increases, and 
asymptotically approaches the standard model value  of $\approx 1$ for very 
large values of $M/\L$. This is because the corresponding kinematic
suppression for high selectron masses makes the GMSB effects progressively
insignificant with respect to the SM effects. However, a large region of the 
parameter space shows an unambiguously measurable asymmetry,
much in the same way as is $\cal{A^{\gamma\gamma}}$ discussed above. 
It should be mentioned that both the graphs are drawn assuming a 
$100 \%$ polarization efficiency of the $e^-$ beam and completely 
unpolarized $e^+$ beam.

The range in the parameter space  that is covered in our rather
conservative approach here includes practically the entire region within the
reach of the Large Hadronic Collider (LHC). If, therefore, it is  expected that
a SUSY signal is observed at the LHC, then one may aspire to resolve the  
uncertainty  regarding the SUSY breaking mechanism by performing experiments
with polarized electrons at a high-energy $e^+e^-$ linear collider.

We end this discussion with a few remarks. We have shown only two cases of NLSP
production here. Situations where non-NLSP superparticles are produced and
subsequently decay into via the NLSP channel to a gravitino can also have
significant asymmetries for exclusive final states. For example, with a 
neutralino NLSP, the production of right selectron pairs can lead to the
asymmetric $e^{+} e^{-} + p_T\slash $  ~signal. Similar signals follow from
pair-produced neutralinos when a right selectron is the NLSP. Also, though we
have not shown any results for $N_m = 2$, this particular case deserves  
closer attention as over a large region of the parameter space, the right
slepton and the lightest neutralino are nearly degenerate, and the resulting
signals can be quite interesting. A detailed study of the above points will
be reported in a subsequent paper.

\newpage
\centerline{\bf Table 1}

\begin{table}[htbp]
\begin{center}
\begin{tabular}{|c|c|c|c|c|c|c|}
\hline
$N_m$&$\mu$&$\tan\beta$&$M$ (TeV)&$\L$ (TeV)&$M_{\chi^0}$ (GeV)&$M_{\tilde
{e_R}}$ (GeV)\\
  &  &  &  &  &  &  \\
\hline
1&300&2&100&50&61&91\\
1&900&2& 50&25&33&54\\
2&500&2&50&25&66&68\\
2&500&20&100&50&139&127\\
2&800&2&50&25&68&68\\
3&100&2&100&50&72&150\\
3&100&20&50&25&62&84\\
3&800&2&500&50&199&154\\
3&900&20&500&100&40&52\\
\hline
\end{tabular}
\end{center}
\end{table}
Some sample mass values for the lightest neutralino and the right selectron
with different GMSB parameters. 

%
%
%
 

\begin{thebibliography}{99}
\bibitem{note1} The mass spectrum depends crucially on the messenger scale
dynamics. Also, for a strongly interacting messenger sector, right sleptons
can naturally become the NLSP; see the first reference in \cite{gmsbsigs}.
\end{thebibliography}
\end{document}